\documentclass[a4paper,12pt]{article}

\usepackage{epsfig}
\usepackage{url}
\usepackage{graphicx,color}
\usepackage[psamsfonts]{amssymb}
\usepackage{amsmath}
\usepackage{indentfirst}
\usepackage{mathrsfs}
\usepackage{booktabs}
\usepackage{tabularx}
\usepackage{authblk}
\usepackage{setspace}
\usepackage{natbib}

\usepackage[margin=2.5cm]{geometry}

\begin{document}

\title{Estimation of the Hawkes Process With Renewal Immigration Using the EM Algorithm}

\author[1]{Spencer Wheatley\thanks{swheatley@ethz.ch}}
\author[1]{Vladimir Filimonov\thanks{vfilimonov@ethz.ch}}
\author[1,2]{Didier Sornette\thanks{dsornette@ethz.ch}}
\affil[1]{Dept. of Management, Technology and Economics, ETH Z\"{u}rich,\newline Z\"{u}rich, Switzerland}
\affil[2]{Swiss Finance Institute, c/o University of Geneva}

\date{\today}

\maketitle


\begin{abstract}

We introduce the \emph{Hawkes process with renewal immigration} and make its statistical estimation possible with two \emph{Expectation Maximization} (EM) algorithms. The standard \emph{Hawkes process} introduces \emph{immigrant points} via a Poisson process, and each immigrant has a subsequent cluster of associated offspring of multiple generations. We generalize the immigration to come from a Renewal process; introducing dependence between neighbouring clusters, and allowing for over/under dispersion in cluster locations. This complicates evaluation of the likelihood since one needs to know which subset of the observed points are immigrants. Two EM algorithms enable estimation here: The first is an extension of an existing algorithm that treats the entire branching structure - which points are immigrants, and which point is the parent of each offspring - as missing data. The second considers only if a point is an immigrant or not as missing data and can be implemented with linear time complexity. Both algorithms are found to be consistent in simulation studies. Further, we show that misspecifying the immigration process introduces signficant bias into model estimation-- especially the \emph{branching ratio}, which quantifies the strength of self excitation. Thus, this extended model provides a valuable alternative model in practice.
 
\end{abstract}

\newpage

\section{Introduction}

The \emph{Hawkes process}~\citep{Hawkes1971_orig,Hawkes1971} is a linearly self-exciting conditional Poisson point process. It can be mapped onto a a branching process in which initial \emph{immigrant events} can produce subsequent \emph{offspring events}. Further, all realized offspring may produce offspring in the same way. Thus each immigrant along with its multi-generational tree of offspring forms a \emph{cluster}. This model has attracted a lot of attention since it parsimoniously combines exogenous activity (immigration) with endogeneous self exciting dynamics (offspring).
\\
The seminal application of the Hawkes model was within seismology~\citep{KKno1981,KKno1987,Ogata1988}, where its spatio-temporal marked extension is being successfully used for modeling so-called triggered seismicity: the self-generated aftershock sequence after individual earthquakes (see for instance review in~\cite{Ogata2013}). Recent applications include: modeling genomic events along DNA~\citep{ReynaudBouret2010}; neural spike trains~\citep{Krumin2010}; brain seizures~{\citep{Sornette2010epileptic}}; and the spread of violence~\citep{LewisMohler2010} and crime~\citep{Mohler2011}. In financial and econometrical applications, the Hawkes process is becoming the gold standard for modeling high frequency fluctuations of financial prices (see for instance~\citep{Bowsher2007,Bauwens2009,FilimonovSornette2012_Reflexivity,Muzy2013hawkes}).
\\
In the theory of Hawkes processes, the immigration process, i.e., the location of clusters, is a Poisson process. This implies that the cluster locations are independent. In this form, the likelihood of the Hawkes model can be evaluated and thus Maximum Likelihood Estimation (MLE) can be performed. An extension of this model with inhomogenous Poisson immigration~\citep{LewisMohler2010} may also be estimated by maximum likelihood. 
\\
Poissonian immigration corresponds to exponentially distributed waiting times between immigrants. Here we introduce a natural extension: the~\emph{Hawkes process with renewal process immigration}. The renewal immigration process, like the Poisson process, has i.i.d (independent and identically distributed) waiting times, but with a general waiting time distribution instead of the exponential distribution. The exponential distribution has a dispersion index value of one (the ratio of the variance to the mean of interevent times), and is thus called equi-dispersed. The renewal process can describe both under- and over- dispersed immigration (dispersion index less than and greater than $1$, respectively). For instance, with a Weibull waiting time distribution, one obtains this flexibility by introducing only one extra parameter relative to the Poisson process. This flexibility comes at the cost of making the evaluation of the likelihood impossible, thus making direct MLE practically impossible.
\\
To the best of our knowledge, there are no existing mathematical papers focused on this Hawkes model with Renewal immigration and more generally for renewal cluster processes and branching process with renewal process immigration. There have been some similar models proposed for applications in climatology but with either very restrictive model assumptions, or less desirable estimation properties. In~\cite{Cowpertwait2000}, a renewal cluster model was proposed for clustering of rainfall events. In this model, clusters are separated, from the end point of the last cluster to the starting point of the next cluster, by a renewal process. Thus no overlap in clusters is allowed, which is a severe assumption in many applications. This simplification makes maximum likelihood estimation easy. Further, using Bartlett-Lewis type clustering~\citep{CoxIsham1980}, only the most recent generation point is fertile in this model. That is, the offspring are distributed after their immigrant in a finite renewal process with random termination size. In~\cite{Salim2003}, a Bartlett-Lewis cluster process with renewal process immigration was considered for clustering of rainfall events. Introducing overdispersion into the immigration process was motivated by the occasional observation of long periods without rainfall. The authors maximized a quasi-likelihood to estimate the model.
\\
In the present paper, we propose a novel way to calibrate such generalized renewal Hawkes processes using the \emph{Expectation Maximization} (EM) algorithm~\citep{Dempster1977}. We show that the problem of missing data (the branching structure), which is necessary for the derivation of the complete data likelihood function, can be easily addressed within the EM approach. The EM algorithm has already been used for the estimation of the standard Hawkes process ~\citep{Marsan2008,Veen2008,LewisMohler2011}. Here we extend this approach to the case of renewal process immigration, and introduce another EM algorithm with a reduced set of missing data.
\\
The structure of the paper is as follows. Section 2 presents the general structure of the standard Hawkes process with Poisson immigration together with the Hawkes process with renewal immigration. Section 3 introduces the complete data EM algorithm for the estimation of Hawkes process with renewal immigration. Section 4 introduces the semi-complete data EM algorithm using a minimal set of missing data. In Section 5, the computation of likelihood and performance of goodness of fit tests for estimates of the Hawkes model with renewal immigration are discussed. Section 6 presents results of Monte Carlo studies of the complete data EM algorithm including starting point selection, non-parametric methods, consistency, model selection, and robustness of branching ratio estimation under immigrant process misspecification. In Section 7, the semi-complete data EM algorithm is shown to be a computationally efficient method for the estimation of the Hawkes process with inhomogeneous Poissonian immigration. In section 8, we conclude with a discussion on the significance and relevance of our findings, and directions for further research.

\section{The Hawkes Process With Renewal Immigration}

Consider a sequence of random event times $\{T_{i}\}_{i\in \mathbb{N}}$, such that $T_i<T_{i+1}$ with \emph{inter-event waiting times} $W_i=T_i-T_{i-1}$. This sequence defines an \emph{univariate point process} with \emph{counting process} $N(t)=\sum_i 1_{t_i\leq t}$. Denote a realization of the point process $\boldsymbol{t}_{1:n}=\{t_{1},\dots,t_{n}\}$ on $(0,r]$ with \emph{stopping time} $r$ where $n=N(r)$. When necessary, we denote the \emph{history} of the process $\mathcal{H}_{t-}=\{t_{1},\dots,t_{i}: t_{i}<t\}$ on time window $(0,t]$. A point process $\{T_{i}\}$ can be defined by its \emph{conditional intensity} $\lambda(t|\mathcal{H}_{t-})$ which is the instantaneous conditional probability of an event occurring~\citep{VereJones2003_vol1}. The \emph{compensator} of the point process is the expected value of the counting process at time $t$: $\Lambda(t|\mathcal{H}_{t-})=\int_{0}^{t}\lambda(s|\mathcal{H}_{s-})ds=\mathrm{E}[N(t)|\mathcal{H}_{t-}]$.
\\
The Hawkes process~\citep{Hawkes1971_orig,Hawkes1971} is a self-exciting conditional Poisson process with the following linear conditional intensity function~\mbox{\citep{Hawkes1971_orig,Hawkes1971}}:
\begin{equation}\label{eq:HawkesCondInt}
	\lambda(t|\mathcal{H}_{t-})= \mu + \Phi(t|\mathcal{H}_{t-}),
\end{equation} 
where $\mu\in(0,\infty)$ is the \emph{background intensity} and the conditional self-exciting term is $\Phi(t|\mathcal{H}_{t-})$ where 
\begin{equation}\label{eq:HawkesPhi}
	\Phi(t|\mathcal{H}_{t-})= \sum_{i:t_{i}<t} \eta h(t-t_i).
\end{equation} 
The function $\eta h(t-t_{N(t)})$ is the intensity of an inhomogeneous Poisson process originating at each observed point, $t_i$. It is also called the \emph{memory kernel}. The function $h(.)$ is the \emph{offspring density}, a pdf (probability density function) only giving mass to positive support. The parameter $\eta$ is the \emph{branching ratio}, a non-negative constant determining the strength of self-excitation.
\\
The Hawkes process can also be considered as a branching process: $\mu$ is the \emph{immigration intensity}, and $\eta h(t-t_{N(t)})$ is the \emph{offspring intensity} -- an inhomogeneous Poisson process triggered by each observed point. The branching ratio $\eta$ is the expected number of offspring of each point, and the offspring density provides the law for the parent-child inter-event times. Due to the autoregressive nature of the process, it becomes non-stationary for $\eta > 1$ (the case
$\eta =1$ is borderline stationary with non-standard scaling properties \citep{SaichevSor14}).
The branching process is constructed as follows: the Poissonian immigration process $\{T_{i}^{(0)}\}$ introduces immigrant points into the zeroeth generation; each immigrant generates a subsequent offspring process, which introduces first-generation \emph{offspring} events $\{T_{i}^{(1)}\}$; each offspring may introduce second-generation offspring $\{T_{i}^{(2)}\}$ in the same way; and so on over many generations. The union of the immigrants and offspring of multiple generations defines the Hawkes process ($\{T_{i}^{(0)}\}\cup\{T_{i}^{(1)}\}\cup\dots$). Further, by summing the mutually independent immigration and offspring intensities, one recovers the conditional intensity~\eqref{eq:HawkesPhi}. A realization of this conditional intensity is in the top panel of Fig.~1. 
\\

\begin{figure}[t!]
 \begin{center} 
\centerline{\includegraphics[width=10cm]{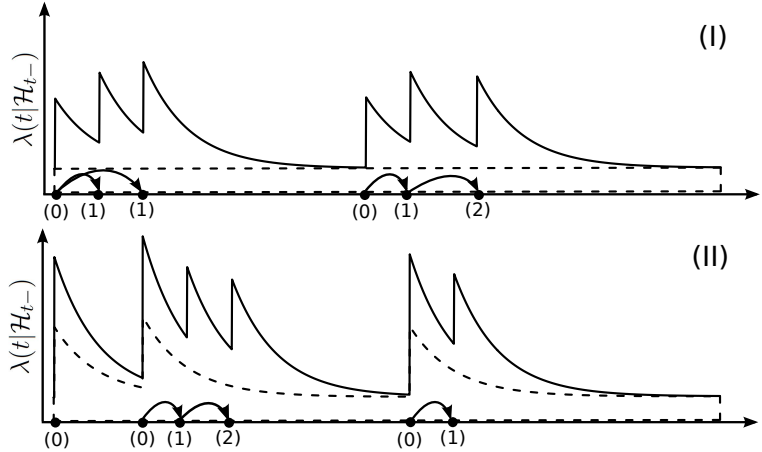}}
\caption{Illustration of the conditional intensity functions for (I) the Hawkes process~\eqref{eq:HawkesCondInt} and (II) the Hawkes process with over-dispersed renewal process immigration~\eqref{eq:HawkesRenewalCondInt}. The dotted line represents the intensity of immigration, and the solid line corresponds to the total intensity. The arrows indicate parenthood and the numbers in parentheses indicate the generation of offspring. }
\end{center}
\label{fig1}
\end{figure}

The standard definition of the Hawkes process~\eqref{eq:HawkesCondInt} assumes constant or time-varying but deterministic immigration intensity $\mu(t)$, or in other words a Poissonian immigration process $\{T_{i}^{(0)}\}$. Here we extend the definition by introducing the \emph{Hawkes Process with Renewal immigration}. For this, we allow the immigration process $\{T_{i}^{(0)}\}$ to be a general Renewal process, where waiting times $W_i>0$ are i.i.d from some pdf $g(w)$ that is not necessary exponential. The conditional intensity of this process is then given by:
\begin{equation}\label{eq:HawkesRenewalCondInt}
	\lambda(t|\mathcal{H}_{t-},I[N(t)]) = \mu(t-t_{I[N(t)]}) + \Phi(t|\mathcal{H}_{t-}),
\end{equation} 
where we have defined the index function for immigrants 
\begin{equation}
I[N(t)]=k\in {1,\dots,N(t)}~,
\label{def_indexmostrecent}
\end{equation}
which provides the point index $k$ of the most recent immigrant event $t_k$ before $t$ in the Hawkes realization $\boldsymbol{t}_{1:n}$. The conditional intensity of offspring $\Phi$ is given by the same equation~\eqref{eq:HawkesPhi}, and the intensity $\mu(w)$ of the renewal immigration process is defined with 
\begin{equation}\label{eq:renewalInt}
	\mu(w)=\frac{ g(w) }{ 1-G(w) },
\end{equation} 
where $G(w)=\int_0^w g(s)ds$ is the CDF (cumulative distribution function) of waiting times $\{W_i\}$. The density $g$ may be expressed using expression~\eqref{eq:intereventDensity} below. Constant intensity then corresponds to a Poisson process with exponentially distributed inter-event times that are equi-dispersed, and a strictly decaying intensity will provide a sub-exponential distribution of inter-event times that are over-dispersed. The intensity $\mu(t)$ integrates to infinity and thus the process does not go extinct. A realization of this process together with renewal immigrantion is presented in the lower panel of Figure~1. 
\\
Where an immigrant together with its associated offspring form a cluster, the Hawkes process is a \emph{Poisson cluster process}, and the Hawkes process with renewal immigration is thus a \emph{Renewal cluster process}. Thus, the model~\eqref{eq:HawkesRenewalCondInt} introduces ``nearest neighbour'' dependence into the location of clusters, allowing for both under- and over-dispersion in the immigration process itself. In some sense, the over-dispersed renewal process, having ``one step memory'', can be considered as the simplest case of a self-exciting process. Thus, the Hawkes process with renewal immigration is a step towards a model with two stages of self excitation.
\\
A challenging aspect of this model is the estimation procedure. The general form of the log-likelihood of a realization $\boldsymbol{t}_{1:n}$ on $(0,r]$ is given by~\cite{VereJones2003_vol1}:
\begin{equation}\label{eq:lik}
	\log \text{L}(\boldsymbol{\theta};\boldsymbol{t}_{1:n})=\sum_{i=1}^{n}\log\lambda(t_{i}|\mathcal{H}_{t-})-\Lambda(r|\mathcal{H}_{t-}),
\end{equation}
where $\boldsymbol{\theta}$ is the parameter vector of the model. Maximizing this log likelihood~\eqref{eq:lik} with respect to $\boldsymbol{\theta}$ has the interpretation of maximizing the intensity at observed points and minimizing intensity where no points are observed. For the standard Hawkes model, the conditional intensity~\eqref{eq:HawkesCondInt} can be evaluated and thus MLE can be performed. For the Hawkes process with renewal immigration, to evaluate the intensity~\eqref{eq:HawkesRenewalCondInt} and thus the likelihood~\eqref{eq:lik}, one needs to know which events are immigrants. Since this information is unobserved (all events --- immigrants and descendants --- are identical), direct MLE using~\eqref{eq:lik} is not possible. In order to account for this missing data, in the following section we will employ an Expectation Maximization (EM) algorithm.
\\
It is worth explaining the structure of the likelihood as the details are relevant in the following sections. The likelihood~\eqref{eq:likelihood},  
\begin{equation}\label{eq:likelihood}
	\text{L}(\boldsymbol{\theta};\boldsymbol{t}_{1:n})=f(\boldsymbol{t}_{1:n};\boldsymbol{\theta})\text{Pr}_{\boldsymbol{\theta}}\left[ N(r)-N(t_{n})=0 | \boldsymbol{t}_{1:n} \right],
\end{equation}
is the product of the \emph{joint inter-event time density} and the probability that no events occur between the last point and the stopping time. The joint inter-event time density within \eqref{eq:likelihood} can be factored into a product of conditional Poisson marginal inter-event time densities,
\begin{equation}\label{eq:intereventDensity}
  \text{f}(t|\mathcal{H}_{t-})=\lambda(t|\mathcal{H}_{t-})\text{exp}\left( -\int_{t_{N(t)}}^{t}\lambda(s|\mathcal{H}_{s-})ds \right),
\end{equation}
which is the conditional probability of observing a point at $t$ times the probability of no points between the previous point $t_{N(t)}$ and $t$. For an inhomogeneous Poisson process, the density \eqref{eq:intereventDensity} becomes unconditional and, for a homogeneous Poisson process, it becomes an exponential density.

\section{Complete Data EM Algorithm for the Hawkes Process with Renewal Immigration} 
\label{sec:EM}

\subsection{General description of EM algorithms for Hawkes processes}

An Expectation Maximization (EM) algorithm, first introduced in~\cite{Dempster1977}, is an iterative algorithm for performing MLE in the presence of missing data that, if known, would simplify the likelihood. Consider the \emph{complete data} likelihood function $L(\boldsymbol{\theta};\boldsymbol{X},\boldsymbol{Z})$ that has an explicit form for known sets of \emph{observed data} $\boldsymbol{X}$ and \emph{missing data} $\boldsymbol{Z}$. Set $\boldsymbol{X}$ together with set $\boldsymbol{Z}$ forms the \emph{complete data}. Maximization of $\log \text{L}(\boldsymbol{\theta};\boldsymbol{X},\boldsymbol{Z})$ with respect to $\boldsymbol{\theta}$ then results in parameter estimates. When $\boldsymbol{Z}$ is unknown, one can account for the missing data probabilistically using the following iterative procedure.
\\
First, one needs to guess the initial parameter estimates $\boldsymbol{\hat\theta}^{[0]}$. Then each $m$'th iteration consists of two steps. Given the estimates of parameters $\boldsymbol{\hat\theta}^{[m]}$ in the \emph{expectation (E) step}, one needs to calculate the expected value of the complete data log-likelihood with respect to the conditional distribution of missing data~$\boldsymbol{Z}$, given the observed data $\boldsymbol{X}$ and current estimates of parameters $\boldsymbol{\hat\theta}^{[m]}$:

\begin{equation}\label{eq:Q_complete_}
	\text{Q}(\boldsymbol{\theta}|\boldsymbol{X},\boldsymbol{\hat\theta}^{[m]})=
	\mathrm{E}_{\boldsymbol{Z}|\boldsymbol{X},\boldsymbol{\hat\theta}^{[m]}}
	\Big[\log \text{L}(\boldsymbol{\theta};\boldsymbol{X},\boldsymbol{Z})\Big].
\end{equation}

The conditional expectation~\eqref{eq:Q_complete_} requires the derivation of the conditional pdf
$f(\boldsymbol{Z}|\boldsymbol{X},\boldsymbol{\hat\theta}^{[m]})$, which is nothing more than the likelihood function for estimation of missing data~$\boldsymbol{Z}$ given~$\boldsymbol{\hat\theta}^{[m]}$ and~$\boldsymbol{X}$.
In the \emph{maximization (M) step}, the expected likelihood~\eqref{eq:Q_complete_} is maximized to obtain new estimates of parameters $\boldsymbol{\hat\theta}^{[m+1]}$:
\begin{equation}\label{eq:M}
	\boldsymbol{\hat\theta}^{[m+1]}=\arg\max_{\boldsymbol{\theta}}Q(\boldsymbol{\theta}|\boldsymbol{X},\boldsymbol{\hat\theta}^{[m]}).
\end{equation}
The algorithm then proceeds by iterating the E and M steps until the parameter estimates $\boldsymbol{\hat\theta}^{[m]}$ stabilize. With each iteration, parameter estimates are guaranteed not to make the observed data likelihood worse~\citep{Dempster1977}.
\\
An EM algorithm for the Hawkes process was identified in \cite{Marsan2008} and formalized in \cite{Veen2008,Mohler2011}. In particular, this EM algorithm has had strong convergence results proven and tested \citep{Mohler2011}, and in practice converges in less than 50 iterations, given sensible initial parameter estimates. Here this algorithm is extended to the case of Renewal process immigration, and an algorithm with a reduced set of missing data is also presented in Section \ref{sec:semiEM}.
\\
For the case of the Hawkes process with renewal immigration,
$\boldsymbol{X}$ is given by $\boldsymbol{t}_{1:n}$, and unobserved data $\boldsymbol{Z}$ is given by the \emph{branching structure} of the process, which contains information of: (i) immigrant events and (ii)  parenthood of offspring events (see Figure~1). 
This branching structure is typically described with the lower-triangular matrix $\boldsymbol{Z}_{n\times n}$ with diagonal elements $Z_{i,i}=1$ if point $t_{i}$ is an immigrant and $Z_{i,i}=0$ if not; and sub-diagonal elements $Z_{i,j}=1,~j<i$ if point $t_{j}$ is direct parent to point $t_{i}$. Since a point can be either an immigrant or an offspring of a single parent, then each row of the matrix has only one unit element, and the rest of the entries are zero.
\\
Consider the joint density $f(\boldsymbol{t}_{1:n},\boldsymbol{Z}_{n\times n})=f(\boldsymbol{t}_{1:n}|\boldsymbol{Z}_{n\times n})f(\boldsymbol{Z}_{n\times n})$. When the branching structure~$\boldsymbol{Z}_{n\times n}$ for the self-excited Hawkes process with renewal immigrants~\eqref{eq:HawkesRenewalCondInt} is known, the conditional density $f(\boldsymbol{t}_{1:n}|\boldsymbol{Z}_{n\times n})$ can be split into a product of marginal inter-event time densities for independent inhomogeneous Poisson sub-processes (i.e., densities of the form~\eqref{eq:intereventDensity}):
\begin{equation}\label{eq:cond_pdf}
f(\boldsymbol{t}_{1:n}|\boldsymbol{Z}_{n\times n})= 
 \prod_{i=1}^{n} \prod_{j={1}}^{i-1} \left[ \mu(t_i-t_j)e^{-\int_{t_j}^{t_i}\mu(s-t_j)ds} \right]^{ Z_{i,i} 1_{\lbrace I[i]=j \rbrace} } 
 \prod_{i=1}^{n} \prod_{j=J[i]}^{i-1}
 \left[ \eta h(t_i-t_j)e^{-\int_{t_j}^{t_i} \eta h(s-t_j)ds} \right]^{ Z_{i,j} }
\end{equation}
The first term in square brackets is the immigrant inter-event time density $g$ introduced in relation with expression~\eqref{eq:renewalInt}) and $I[i]$ is defined in~\eqref{def_indexmostrecent}. When a lag $t_i - t_j, j<i=1,\dots,n$ is an immigrant inter-event time (i.e., $Z_{i,i} 1_{\lbrace I[i]=j \rbrace}=1$) then $g$ is evaluated at that lag. The second term in the square brackets is the \emph{offspring intevent time density}. When a lag $t_i - t_j, j<i=1,\dots,n$ is a parent-child inter-event time (i.e., $Z_{i,j}=1$) then the offspring inter-event time density is evaluated at that lag. To avoid undefined values of~\eqref{eq:cond_pdf}, the offspring inter-event time density is only evaluated at lags within the support of the offspring density $h$. This is done by defining this index function 
\begin{equation}
J[i]:=\min(j\in \lbrace 1, \dots, i-1 \rbrace~:~h(t_i-t_j)>0)
\label{defindexfun}
\end{equation}
that takes the index $i=1,\dots,n$ of point $t_i$ and returns the index of the most distant previous point $t_j$ with $t_i$ in the support of its offspring density $h(t-t_{j})$. This issue is not present for $g$ since the immigration intensity never vanishes.
\\
Following~\eqref{eq:likelihood}, the complete data likelihood $L(\boldsymbol{\theta} ; \boldsymbol{t}_{1:n},\boldsymbol{Z}_{n\times n})$ is constructed as a product of the joint pdf of observed events $f(\boldsymbol{t}_{1:n},\boldsymbol{Z}_{n\times n})$ and a compensator term which accounts for the probability of observing no event after the last event in each independent subprocess. Thus, after substituting~\eqref{eq:cond_pdf} into ~\eqref{eq:likelihood} and rearranging, the complete data log-likelihood of the Hawkes process with renewal immigration is written as:

\begin{multline}\label{eq:completelogLik}
	\log \text{L}(\boldsymbol{\theta} ; \boldsymbol{t}_{1:n} , \boldsymbol{Z}_{n\times n}) =
	\log \text{f}(\boldsymbol{Z}_{n\times n})+
   \left[\sum_{i=1}^{n} \sum_{j=J[i]}^{i-1} Z_{i,j} \log \eta h(t_{i}-t_{j}) 
   	- \int_{0}^{r}\Phi(s|\mathcal{H}_{s-})ds \right] \\
+ \left[\sum_{i=1}^{n} \sum_{j=1}^{i-1} Z_{i,i} 1_{\lbrace I[i]=j \rbrace} \log\mu(t_{i}-t_{j}) 
 -\sum_{i=1}^{n+1}\sum_{j=1}^{i-1} 1_{\lbrace I[i]=j \rbrace}\int_{t_{j}}^{t_{i}}\mu(s-t_{j})ds \right].
\end{multline}

For compact notation in~\eqref{eq:completelogLik}, we have also denoted $t_{0}=0$ as the starting time and $t_{n+1}=r$ as the stopping time. Neither of these points are included in the sample. As it is seen, the complete data log likelihood is decoupled into a sum of independent terms for offspring and immigrant processes (in square brackets). Thus for a given branching structure $\boldsymbol{Z}_{n\times n}$, the estimation of parameters $\boldsymbol{\theta}$ for the model amounts to independent estimation from i.i.d (independent and identically distributed) samples of immigrant inter-event times and parent-child inter-event times respectively. 
\\
The following subsections describe the E step and M step for the corresponding EM-algorithm that accounts for the unobserved branching structure $\boldsymbol{Z}_{n\times n}$.

\subsection{E Step}

The E step involves evaluating the Q function~\eqref{eq:Q_complete}, which is the expected value of the complete data log-likelihood~\eqref{eq:completelogLik} given the observed data $\boldsymbol{t}_{1:n}$ and previous estimates of parameters $\boldsymbol{\hat\theta}^{[m]}$:

\begin{eqnarray}\label{eq:Q_complete}
	Q(\boldsymbol{\theta}|\boldsymbol{t}_{1:n},\boldsymbol{\hat\theta}^{[m]}) &=&
	\mathrm{E}_{\boldsymbol{Z}_{n\times n}|\boldsymbol{t}_{1:n},\boldsymbol{\hat\theta}^{[m]}}
	\Big[\log \text{L}(\boldsymbol{\theta};\boldsymbol{t}_{1:n},\boldsymbol{Z}_{n\times n})\Big]\propto
	\nonumber \\
	&& \left[\sum_{i=1}^{n} \sum_{j=J[i]}^{i-1} \mbox{Pr}[Z_{i,j}=1|\boldsymbol{t}_{1:i},\widehat{\boldsymbol{\theta}}^{[m]}] \log\eta h(t_{i}-t_{j}) - \int_{0}^{r}\Phi(s|\mathcal{H}_{s-})ds \right]\nonumber \\
	&& + \left[\sum_{i=1}^{n} \sum_{j=1}^{i-1} \mbox{Pr}[ Z_{i,i} 1_{\lbrace I[i]=j \rbrace}=1 |\boldsymbol{t}_{1:i},\widehat{\boldsymbol{\theta}}^{[m]} ]\log\mu(t_{i}-t_{j}) 
\right.\nonumber \\
	&& \left.-\sum_{i=1}^{n+1}\sum_{j=1}^{i-1} \mbox{Pr}[I[i]=j|\boldsymbol{t}_{1:i},\widehat{\boldsymbol{\theta}}^{[m]}]\int_{t_{j}}^{t_{i}}\mu(s-t_{j})ds \right],
\end{eqnarray}
where we have omitted the expectation of the first term in~\eqref{eq:completelogLik}, which is constant with respect to  $\boldsymbol{\theta}$ and is thus irrelevant to the determination of the parameters $\boldsymbol{\theta}$.
\\
To compute~\eqref{eq:Q_complete}, the missing data $\boldsymbol{Z}_{n\times n}$ should be defined probabilistically. We denote the probability weights as 
\begin{equation}
\pi_{i,j}=\mbox{Pr}(Z_{i,j}=1|\boldsymbol{t}_{1:i})
\label{defprobwes}
\end{equation}
and introduce the abbreviation $\pi_{i}=\pi_{i,i}$ for immigrant probabilities. By definition, the weights sums to one: $\sum_{j=1}^{i}\pi_{i,j}=1$, $i=1,..,n$. 
The first event ($i=1$) has $\pi_1=\pi_{1,1}=1$ and is thus an immigrant. The second event ($i=2$) has $\pi_{2,2}+\pi_{2,1}=1$ and thus can either be an immigrant or an offspring with the respective probabilities. 
Each next event has one more parameter in the probability distribution than its predecessor. All these probabilities for the $n$ observed points can be presented as a lower-triangular matrix $\boldsymbol{\Pi}_{n \times n}$ that is, at each iteration of the EM algorithm, equal to the expected value of the branching structure matrix: 
\begin{equation}
\boldsymbol{\Pi}^{[m]}_{n \times n}=\mathrm{E}[ \boldsymbol{Z}_{n\times n} | \boldsymbol{t}_{1:n},\boldsymbol{\hat\theta}^{[m]}]~.
\label{defbranaverastr}
\end{equation}
Finally, we denote conditional probability weights: $\pi_{i,j|k}=\mbox{Pr}[ Z_{i,j}=1 | t_{1:i},I[i]=k ],~j\leq i$ that are abbreviated $\pi_{i|k}=\pi_{i,i|k}$ for immigrants. 
In this notation, probabilities in~\eqref{eq:Q_complete} can be written in the form:
\begin{eqnarray} \label{eq:Q_weights}
	&& \mbox{Pr}[ Z_{i,j}=1|\boldsymbol{t}_{1:i} ] = \pi_{i,j} \nonumber \\
	&& \mbox{Pr}[ I[i]=j |\boldsymbol{t}_{1:i} ] :=\omega_{i,j}
	= \pi_{j}\bar{\pi}_{j+1|j}\dots \bar{\pi}_{i-1|j} \nonumber \\
	&& \mbox{Pr}[ Z_{i,i} 1_{\lbrace I[i]=j \rbrace}=1 |\boldsymbol{t}_{1:i} ] = \mbox{Pr}[ I[i]=j |\boldsymbol{t}_{1:i} ]\pi_{i|j}=\omega_{i,j}\pi_{i|j},
\end{eqnarray}
where we have introduced weights $\omega_{i,k}$ and the bar denotes the complementary probability: $\bar\pi_{i,j|k}=1-\pi_{i,j|k}$. The first line of (\ref{eq:Q_weights})
is just the definition (\ref{defprobwes}). The second line defines the probability
that $j$ is the last immigrant in the series of $i$ events up to time $t_{i}$ as the product of the probability $\pi_{j}$ that $j$ is an immigrant times the probabilities that all following events are not immigrants (conditional on j being an immigrant). The third line defines the probability
that $j$ is the last immigrant before immigrant $i$ in the series of $i$ events up to time $t_{i}$.
\\
To derive the probability weights $\pi_{i,j}$ defined by (\ref{defprobwes}),
we will exploit the branching structure of the Hawkes process with Renewal immigration~\eqref{eq:HawkesRenewalCondInt}, which consists of a superposition of independent subprocesses. According to the thinning property~\citep{VereJones2003_vol1} (that was originally exploited for a similar purpose of ``stochastic declustering'' in~\cite{Zhuang2002}), the probability that an observed event $t_i$ comes from one of the subprocesses is equal to the proportion of the subprocess' conditional intensity at $t_i$ in the total conditional intensity at the same time $t_i$.
\\
Conditional probability weights $\pi_{i,j|k}$ can be calculated using the \emph{complete data conditional intensity}~\eqref{eq:HawkesRenewalCondInt}, where the immigrant events $\{t_{i}^{(0)}\}$ are known. To derive unconditional probability weights $\pi_{i,j}$, one needs to introduce the \emph{incomplete data conditional intensity}:
\begin{equation}\label{eq:incompl_lambda}
	\lambda_{*}(t_i|\mathcal{H}_{t_{i-}}) = \mu_{*}(t_i|\mathcal{H}_{t_{i-}}) +\Phi(t_i|\mathcal{H}_{t_{i-}}),
\end{equation}
where the incomplete data conditional intensity of immigration $\mu_{*}(t_i|\mathcal{H}_{t_{i-}})$ is a weighted mixture of immigrant intensities
\begin{equation}\label{eq:incompl_mu}
 	\mu_{*}(t|\mathcal{H}_{t_{i-}})=\sum_{j=1}^{N(t)} \omega_{N(t),j}\cdot  \mu(t-t_{j}), 
\end{equation}
with weights $\omega_{N(t),j}$~\eqref{eq:Q_weights} equal to the probability that the event $j$ at time $t_j$ is the most recent immigrant before $t_{N(t)}$.
\\
Finally, the estimation of probability weights for given observed data $\boldsymbol{t}_{1:n}$ and parameters $\boldsymbol{\hat\theta}^{[m]}$ can be written in the following form:
\begin{eqnarray}\label{eq:probs}
	\pi_{i}=\frac{\mu_{*}(t_{i}|\mathcal{H}_{t_{i-}})}
	{\mu_{*}(t_{i}|\mathcal{H}_{t_{i-}})+\Phi(t_{i}|\mathcal{H}_{t_{i-}})},
	&~&\pi_{i|k}=\frac{\mu(t_{i}-t_{k})}
	{\mu(t_{i}-t_{k})+\Phi(t_{i}|\mathcal{H}_{t_{i-}})},\quad k<i=2,...,N 
	\nonumber  \\
	\pi_{i,j}=\frac{\eta h(t_{i}-t_{j})}
	{\mu_{*}(t_{i}|\mathcal{H}_{t_{i-}})+\Phi(t_{i}|\mathcal{H}_{t_{i-}})},
	&~&\pi_{i,j|k}=\frac{\eta h(t_{i}-t_{j})}
	{\mu(t_{i}-t_{k})+\Phi(t_{i}|\mathcal{H}_{t_{i-}})},\quad j,k<i,~i=2,...,N
\end{eqnarray}
Probability weights $\pi$ and $\omega$, which enter the Q function~\eqref{eq:Q_complete}, can be jointly computed in a recursive way, iterating over all observed events. For each event $i$ at time $t_i$, we denote the probability weight vectors $\boldsymbol{\pi}_{i}=(\pi_{i,1},\dots,\pi_{i,i})$ and $\boldsymbol{\omega}_{i}=\left(\omega_{i,1},...,\omega_{i,i-1}\right)$. The first event is set to be an immigrant ($\pi_{1,1}=1,~\omega_{2,1}=1$). Looking at the weight vector makes the recursive relation clear:
\begin{eqnarray}\label{eq:recursive}
\boldsymbol{\omega}_{i}&=&\left(
\pi_{1}\bar\pi_{2|1}\dots\bar\pi_{i-1|1} ~ , ~ \dots ~ , ~ \pi_{j}\bar\pi_{j+1|j}\dots\bar\pi_{i-1|j}~ , ~ \dots ~ , \pi_{i-1}
\right)
	\nonumber  \\
&=& \left( \left( \boldsymbol{\omega}_{i-1} \circ (\bar\pi_{i-1|1},...,\bar\pi_{i-1|j},...,\bar\pi_{i-1|i-2}) \right)~,~\pi_{i-1} \right)
\end{eqnarray}
This recursive equation~\eqref{eq:recursive} expresses that the weight vector $\boldsymbol{\omega}_{i}$ is the Hadamard product (e.g., $(a,b)\circ(c,d)=(ac,bd)$) of the previous weight vector $\boldsymbol{\omega}_{i-1}$ and a vector of complement probabilities; and with $\pi_{i-1}$ concatenated on the end of the product. That is, taking the weight vector $\boldsymbol{\omega}_{i}$, discarding the last element from the vector, and then removing the last $\pi$ probability weight from each remaining element of the vector, one obtains the previous weight vector $\boldsymbol{\omega}_{i-1}$. Having the weight vector $\boldsymbol{\omega}_{i-1}$ then makes it possible to compute the necessary $\pi$ probablility weights (using~\eqref{eq:probs}) to compute the next weight vector $\boldsymbol{\omega}_{i}$. This iteration is done for $i=2,\dots, n$, producing all necessary probability weights for the E step.

\subsection{M Step}

After the estimates of probability weights $\pi$ and $\omega$ for the given estimates of parameters $\boldsymbol{\hat\theta}^{[m]}$ and observed data $\boldsymbol{t}_{1:n}$ are obtained, 
one can maximize the expected complete data log-likelihood (Q function)~\eqref{eq:Q_complete} with respect to the parameter vector $\boldsymbol{\theta}$ to obtain new estimates $\boldsymbol{\hat\theta}^{[m+1]}$. However rather than perform multivariate optimization of~\eqref{eq:Q_complete}, one can exploit the decoupling of the immigrant and offspring terms in~\eqref{eq:Q_complete}.
Decomposition of the Q function~\eqref{eq:Q_complete} into the expected log-likelihood of the immigration process intensity $\mu(.)$ and the offspring process intensity $\eta h(.)$ allows one to estimate the two processes independently when they do not share any common parameters.
\\
By recalling the form of an inter-event time density for an inhomogeneous Poisson process \eqref{eq:intereventDensity}, it can be seen that the M step estimation of the immigration process intensity $\mu(.)$ -- i.e., maximization of the expected complete data log-likelihood presented in the second square brackets in~\eqref{eq:Q_complete} -- is simply the maximization of the expected complete data log-likelihood for the inter-event times $t_i^{(0)}-t_{i-1}^{(0)}$ of the inter-event time density $g(w;\boldsymbol{\theta}_g)$ with parameter vector $\boldsymbol{\theta}_g,$
\begin{equation}\label{eq:gMstep}
	\boldsymbol{\hat\theta}_g
	 =\arg \max_{\boldsymbol{\theta}_g} \sum_{i=1}^{n}\sum_{j=1}^{i-1}
	 \omega_{i,j}\pi_{i|j}\log g(t_i-t_j;\boldsymbol{\theta}_g),
\end{equation}
where the weights $\omega_{i,j}\pi_{i|j}$ denote the probability $\mbox{Pr}[ Z_{i,i} 1_{\{ I[i]=j\}}=1 |\boldsymbol{t}_{1:i} ]$ that $t_j$ and $t_i$ are immigrants with no other immigrant events between them
as defined by expression~\eqref{eq:Q_weights}. In other words, expression~\eqref{eq:gMstep} is the maximum likelihood density estimation with a weighted i.i.d. sample. With the determination of the
estimate $\widehat{g}$, the immigrant intensity can then be computed using~\eqref{eq:renewalInt}. A non-parametric estimate of the density is possible, but numerical stability issues arise when computing the estimate for $\mu(.)$ as the denominator becomes very small.
\\
Estimation of the offspring intensity, requiring maximization of the expected complete data log-likelihood presented in the first brackets in~\eqref{eq:Q_complete}, is perfomed by separately estimating the branching ratio $\eta$ and the offspring density $h(.)$.
\\
The explicit MLE for the branching ratio parameter~$\eta$ can be obtained by analytically maximizing \eqref{eq:Q_complete} with respect to $\eta$:

\begin{equation}\label{eq:etaEM}
\widehat{ \eta }=\frac{ \sum_{i=1}^{n}\sum_{j=1}^{i-1} \widehat{\pi}_{i,j} }{ \sum_{i=1}^{n}\widehat{H}(r-t_{i}) }=\frac{ n- \sum_{i=1}^{n} \widehat{\pi}_{i} }{ \sum_{i=1}^{n}\widehat{H}(r-t_{i}) }~,
\end{equation}
where $H$ is the CDF: $H(t)=\int_0^t h(s)ds$.
The estimated branching ratio~\eqref{eq:etaEM} is the ratio of the expected number of offspring to a number, which is slightly smaller than the total number $n$. The denominator inflates the estimate 
of the number of offspring in the finite observation window to account for unobserved offspring expected to occur after the stopping time $r$ \citep{SaiSorintervent1,SaiSorintervent2,SaiSorintervent3}.
\\
Estimation of the offspring density $h(t;\boldsymbol{\theta}_h)$, parameterized by $\boldsymbol{\theta}_h$, is density MLE with an i.i.d sample with sample weights,
\begin{equation}\label{eq:hMstep}
	\boldsymbol{\hat\theta}_h
	 =\arg \max_{\boldsymbol{\theta}_h} \sum_{i=1}^{n}\sum_{j=J[i]}^{i-1}
	 \pi_{i,j}\log \text{h}(t_i-t_j;\boldsymbol{\theta}_h),
\end{equation}
where the sample weights $\pi_{i,j}$ are the probability that $t_j$ is parent to $t_i$~\eqref{eq:probs}. Non-parametric estimation of the offspring density $h(.)$ is straight-foward, for example assuming it to be a piecewise-constant function on intervals of length $\Delta$. The estimator is then given by
\begin{equation}\label{eq:histEM}
	\widehat{h}(t)=\frac{1}{n \Delta }\sum_{i=1}^{n}\sum_{j=1}^{i-1} \pi_{i,j}
	1_{t_{i}-t_{j}\in(t-\Delta /2,t+\Delta /2)}.
\end{equation}

\subsection{Computational Efficiency \& Approximations }\label{sec:compEff}

In the proposed EM algorithm, one needs to store $O(n^2)$ probability weights and inter-event times
for the estimation of the offspring density from i.i.d data, as shown from expression \eqref{eq:hMstep}. Estimation of the immigration density~\eqref{eq:gMstep} and the branching ratio~\eqref{eq:etaEM} only require the immigrant probabilities, i.e., $O(n)$ probability weights and inter-event times. One approach to address the quadratic complexity of the estimation of $h(.)$ is to consider a reduced set of missing data. This is done in the following section. However, within the current algorithm, there are two approaches that may be taken to speed up the estimation of $h(.)$.
\\
 The first one is a Monte Carlo approach, which works as follows. Choose an ``effective sample size'', e.g., the expected number of offspring points $n_h =\lfloor n - \sum_{i=1}^{n} \pi_i \rfloor$, take a sample (with replacement) of size $n_h$ from all positive inter-event times $t_i-t_j, j<i=1,\dots,n$ where the probability of selecting an inter-event time is proportional to its corresponding sample weight $\pi_{i,j}$. Then estimate $h(.)$ on this unweighted sample. In~\eqref{eq:histEM} and other simple non-parametric density estimation techniques, the inclusion of sample weights is easy.  However, this approach also allows for other non-parametric techniques where considering sample weights is not as natural.
 Thus, more complicated and potentially less suitable non-parametric estimation methods -- for example choosing a smoothness penalty for the estimate and solving for the estimate using variational calculus as in \cite{LewisMohler2010} -- should not be necessary. 
\\
The second approach assumes that the offspring density has finite memory, i.e., a finite support with upper endpoint $t_f$. Then, take  $n_f =\max(\lbrace N(t_i)-N(t_j) :~t_i-t_j<t_f) \rbrace)$ to be the largest number of points observed within the support of the density. Thus, the E step and M step only need to be performed on lags $t_i-t_j,~i=1,\dots,n,~j=\max (1,i-n_f),\dots,i-1$. This reduces the memory requirements and computational complexity from $O(n^2)$ to $O(n n_f)$ with $n_f\leq n$ without introducing error into the procedure. Taking $n_f$ too small will bias downward the estimation of the branching ratio -- this is clear from~\eqref{eq:etaEM} where the denominator will be too large. However, for Hawkes processes with light tailed offspring distributions, this may provide a good approximation. Further, $n_f$ may be adaptively chosen at each iteration of the EM algorithm as one obtains an idea on which support most of the mass is distributed.

\section{The Semi-Complete-Data EM Algorithm }\label{sec:semiEM}

In this section, an alternative EM algorithm with a reduced set of missing data is proposed. It may be used for more computationally efficient estimation of the Hawkes process with renewal process immigration. It may also be used to estimate the Hawkes process with deterministic inhomogeneous Poissonian immigration intensity $\mu(t)$ -- the most efficient implementation allowing this to be done in linear time (see Sec. \ref{sec:study2}).
\\
Instead of the missing data being the entire branching matrix, here it is reduced to only the diagonal elements $\mathrm{diag}(\boldsymbol{Z}_{n\times n})=\lbrace Z_{i,i} \rbrace_{i=1,\dots,n}$; i.e., it is 
reduced to the indicator variables for if a point is an immigrant or not. This is abbreviated by $\boldsymbol{Z}_{1:n}$ and called the \emph{immigrant vector}. Thus, given the \emph{semi-complete data} $\lbrace \boldsymbol{t}_{1:n},\boldsymbol{Z}_{1:n} \rbrace$, the \emph{semi-complete data log-likelihood}~\eqref{eq:semiCompletelogLik} may be written:
\begin{eqnarray}\label{eq:semiCompletelogLik}
&& \log \text{L}(\boldsymbol{\theta}; \boldsymbol{t}_{1:n},\boldsymbol{Z}_{1:n} ) \propto \nonumber \\
&& \sum_{i=1}^{n} (1-Z_{i}) \log \Phi(t_{i}|\mathcal{H}_{t_{i-}}) - \int_{0}^{r}\Phi(s|\mathcal{H}_{s-})ds \nonumber \\
&& + \sum_{i=1}^{n}\sum_{j=1}^{i-1} Z_{i} 1_{\lbrace I[i]=j \rbrace} \log \mu(t_{i}-t_{j}) - \sum_{i=1}^{n+1}\sum_{j=1}^{i-1} 1_{\lbrace I[i]=j \rbrace}\int_{t_{j}}^{t_{i}}\mu(s-t_{j})ds
\end{eqnarray}

The derivation is similar to that of the complete data log-likelihood~\eqref{eq:completelogLik}. Here immigration and offspring processes are separated, but the offspring processes are not decoupled from each-other, unlike in~\eqref{eq:completelogLik} where they are decoupled. The Q function follows by taking the expectation of~\eqref{eq:semiCompletelogLik} with respect to $\boldsymbol{Z}_{1:n}$ given the observations and parameter values. The full expression is omitted for brevity (its structure is already known from~\eqref{eq:semiCompletelogLik}). Because the parts of~\eqref{eq:completelogLik} and~\eqref{eq:semiCompletelogLik} concerning immigration are identical, the immigration part of $Q$ is identical to the immigration part in~\eqref{eq:Q_complete}. The offspring part will be discussed shortly. For the E-step, the weights are computed using~\eqref{eq:probs}, except that only the immigrant-specific probabilities $\pi_i$ and $p_{i|k}$ are needed. Thus the E-step memory requirements here are $O(n)$ rather than $O(n^2)$ as in the complete data case. Regarding the M-step, estimation of $\mu(.)$ will be the same as in the complete-data case~\eqref{eq:gMstep}. The offspring density $h(.)$ and the branching ratio $\eta$ will be jointly estimated by numerically maximizing the part of the Q function concerning the memory kernel:
\begin{equation}\label{eq:semiM}
(\widehat{\boldsymbol{\theta}}_{h},\widehat{\eta})=\arg\max_{(\boldsymbol{\theta}_{h},\eta)} \sum_{i=1}^{n}
(1-\pi_{i}) \log\left( \sum_{j:t_{j}<t}\eta h(t_i-t_j;\boldsymbol{\theta}_{h}) \right) - \int_{0}^{r} \sum_{j:t_{j}<s}\eta h(s-t_j;\boldsymbol{\theta}_{h})ds. 
\end{equation}

While this method is very useful for parametric estimation (see Sec. \ref{sec:study2}), the non-parametric estimation of $h(.)$ will be difficult, due to the fact that, for example, unit mass and positivity must be enforced. Moreover, the offspring density must be evaluated at $O(n^2)$ lags $t_i-t_j$. The Monte Carlo approach proposed in Sec. \ref{sec:compEff} cannot reduce the computation here since $\pi_{i,j},~j<i$ are unknown. However, the second trick of assuming finite support of $h(.)$ can reduce the computation to $O(n n_{h})$. Further, using an exponential offspring density (or any linear combination of them), a recursive relationship \citep{Ozaki1979} reduces the complexity to $O(n)$.

\section{Inference and Goodness of Fit}\label{sec:GoF}

\subsection{Computation of likelihoods and p-values}

Computing likelihoods and performing goodness of fit tests to obtain p-values for estimates of the Hawkes model with Renewal immigration requires evaluating the complete data conditional intensity~\eqref{eq:HawkesRenewalCondInt}. 
Thus the immigrant vector $\boldsymbol{Z}_{1:n}=\lbrace Z_{i,i} \rbrace_{i=1,\dots,n} \in \lbrace0,1\rbrace^{n} $ must be known.  
\\
As will be shown, likelihoods and p-values must be computed for each immigrant vector, and aggregated. In general, there are $2^{n-1}$ possible valid immigrant vectors (the first point is set to be an immigrant). Immigrant vectors are indexed as $\boldsymbol{z}^{(i)}_{1:n},~i=1,\dots,2^{n-1}$, where the index i is one more than the decimal representation of the binary number in the immigrant vector excluding the first element, i.e., $\boldsymbol{z}^{(1)}_{1:n}=(1,0,\dots0,0)$ ,$\boldsymbol{z}^{(2)}_{1:n}=(1,0,\dots0,1)$, $\boldsymbol{z}^{(3)}_{1:n}=(1,0, \dots,1,0)$, etc. To simplify computation, a Monte Carlo approach can be used where sample averages of likelihoods and p-values are taken. Specifically, given the probabilistic description of the branching structure obtained in the E step (probability weights~\eqref{eq:Q_weights}), an ensemble of realizations of the immigrant vector may be generated, and likelihoods and p-values computed for each realization. Ensemble average likelihoods and p-values may then be taken. A Monte Carlo study of the inferential power of these statistics is conducted in Sec. \ref{sec:caseComplete}.

\subsection{Simulating the Immigrant Vector}

One needs to simulate realizations $\boldsymbol{z}_{1:n}=(z_1,\dots,z_n)$ of the random variable $\boldsymbol{Z}_{1:n}$ with binary state space of dimension $n$. For this, an acceptance-rejection thinning type algorithm \citep{Lewis1979} is used : 
Initialize by setting $\boldsymbol{z}_{1:n}=(1,0,\dots,0)$ since the first point is treated as an immigrant. Next, for each following event $t_i$, $i=2,\dots, n$, Bernoulli random variables with probabilities $\pi_{i|1}$~\eqref{eq:probs} are generated, and the first success is taken (at $i=k$) as the second immigrant (so $z_{k}=1$); then the third immigrant is selected in the same way with probabilities $\pi_{i|k},~i=k+1,\dots,n$; and so on. This thinning procedure is repeated until the stopping time $r$ is reached, resulting in a realization $\boldsymbol{z}_{1:n}=\boldsymbol{z}_{1:n}^{(a)}$ where $a\in \lbrace 1,\dots,2^{n-1} \rbrace$. This procedure can be repeated $l$ times, and the sample set $\lbrace a_i \rbrace_{i=1,\dots,l}$ contains the (possibly repeating) indices of the  sampled immigrant vectors.

\subsection{Likelihood}

We now present a way to calculate a likelihood value for the Hawkes model with renewal immigration. This procedure is valid for both the complete and semi-complete data EM algorithms.  The goal is to have a function by which Hawkes models with and without renewal immigration may be compared in an objective statistical way.  To accomplish this, we treat the renewal process immigration as an inhomogeneous Poisson process with (deterministic) intensity,
\begin{equation}\label{eq:renewalInhom}
  \mu^{(j)}(t)=\mu(t-t_{I[N(t)]}|\boldsymbol{z}_{1:n}^{(j)}), 
\end{equation}
which can be evaluated knowing the immigrant index vector $\boldsymbol{z}_{1:n}^{(j)}$. Thus, plugging in the Hawkes conditional intensity~\eqref{eq:HawkesCondInt} with inhomogeneous Poisson immigration intensity~\eqref{eq:renewalInhom} into the log-likelihood equation~\eqref{eq:lik}, and then transforming to a likelihood, one obtains the \emph{conditional incomplete data likelihood}
\begin{equation}\label{eq:jLik}
   \text{L}(\boldsymbol{\theta};\boldsymbol{t}_{1:n}|\boldsymbol{z}_{1:n}^{(j)})  =
 \prod_{i=1}^{n} \left( \mu^{(j)}(t_i)+ \Phi(t_i|\mathcal{H}_{t_{i-}}) \right)
 \text{exp}\left(-\int_{0}^{r}\mu^{(j)}(s)+\Phi(s|\mathcal{H}_{s-})ds\right).
\end{equation}
for the immigrant vector j. By conditioning, the incomplete data likelihood may then be written,
\begin{equation}\label{eq:LikAgg}
  \text{L}(\boldsymbol{\theta};\boldsymbol{t}_{1:n})  =
 \sum_{j=1}^{2^{n-1}}\text{L}( \boldsymbol{\theta};\boldsymbol{t}_{1:n}|\boldsymbol{z}_{1:n}^{(j)} ) 
 \mathrm{Pr}[ \boldsymbol{Z}_{1:n}=\boldsymbol{z}_{1:n}^{(j)} | \boldsymbol{\theta} ],
\end{equation}
which is a weighted sum of the conditional incomplete data likelihood~\eqref{eq:jLik}. The weighting probabilities in~\eqref{eq:LikAgg} may be computed by probabilities from the E step~\eqref{eq:Q_weights}. However, this is computationally burdensome. Instead, a Monte Carlo approximation of the likelihood~\eqref{eq:LikAgg} ,
\begin{equation}\label{eq:LikAggMC}
  \text{L}(\boldsymbol{\theta};\boldsymbol{t}_{1:n}) \approx \frac{1}{l}\sum_{i=1}^{l} \text{L}\left( \boldsymbol{\theta} ; \boldsymbol{t}_{1:n} | \boldsymbol{z}_{1:n}^{(a_i)} \right) ,
\end{equation}
may be done with sampled immigrant vector indices $\lbrace a_i \rbrace_{i=1,\dots,l}$. The approximate log-likelihood may be obtained by taking the logarithm of this average~\eqref{eq:LikAggMC}. One must be careful with the implementation of these calculations as numerical precision issues may be encountered in the averaging. The logarithm of the incomplete data likelihood~\eqref{eq:LikAgg} or its approximation~\eqref{eq:LikAggMC} may be directly compared with the log-likelihood of the standard Hawkes process~\eqref{eq:lik}. 

\subsection{$p$-Values}

To perform a hypothesis test for an estimated point process model, one often does \emph{residual analysis} \citep{Ogata1988} based on the \emph{time change property} \cite{Papangelou1972}: for point process $\lbrace T_{i} \rbrace_{i \in \mathbb{N}}$ with compensator $\Lambda(t|\mathcal{H}_{t-})$, the set of \emph{transformed times} $\lbrace \tilde{T_i}\rbrace_{i \in \mathbb{N}}$, $\tilde{T_i}=\Lambda(T_{i}|\mathcal{H}_{t-})$, are generated by a unit rate Poisson process. Thus for an observed realization $\boldsymbol{t}_{1:n}$, one can estimate its conditional intensity, transform it to $\tilde{\boldsymbol{t}}_{1:n}$ and test if the resultant process is unit Poissonian.  For instance, one can test if the transformed inter-event times are standard exponential distributed. A popular Portmanteau-type test for this is the Kolmogorov Smirnoff (KS) test \citep{Massey1951}. This test measures the \emph{KS distance} between the empirical transformed inter-event time distribution and the standard exponential null distribution. Under the null hypothesis, the distribution of the KS distance is known and thus a p-value may be computed.
\\
More generally, define the \emph{test statistic} (e.g., the KS distance) as a random variable $S:=S(\boldsymbol{T}_{1:n},\boldsymbol{Z}_{1:n})$, which, under the null hypothesis, has known \emph{reference distribution} $F_0$. Here the observed test statistic $S(\boldsymbol{t}_{1:n},\boldsymbol{z}^{(j)}_{1:n})$ transforms a realization of points $\boldsymbol{t}_{1:n}$, given the immigrant vector for those points $\boldsymbol{z}^{(j)}_{1:n}$. For semi-complete data sets $\lbrace \boldsymbol{t}_{1:n},\boldsymbol{z}^{(j)}_{1:n} \rbrace$, the null hypothesis $H_0^{(j)}$ is an event where the model is true for $\boldsymbol{Z}_{1:n}=\boldsymbol{z}^{(j)}_{1:n}$. Then the \emph{semi-complete data p-values} are
\begin{equation}\label{eq:p_i}
  p^{(j)}=\mathrm{Pr}[S > S( \boldsymbol{t}_{1:n}, \boldsymbol{Z}_{1:n} ) | H_0^{(j)} ]=1-F_{0}(S(\boldsymbol{t}_{1:n},\boldsymbol{z}^{(j)}_{1:n})) ,~j=1,\dots,2^{n-1}.
\end{equation}
  
For the incomplete data set $\lbrace \boldsymbol{t}_{1:n} \rbrace$, the null hypothesis $H_0$ is that the model is true. The test statistic for this, $S(\boldsymbol{t}_{1:n},\boldsymbol{Z}_{1:n})$, is unknown because the immigrant vector is unknown. Thus, by conditioning, the \emph{incomplete data p-value} is
\begin{equation}\label{eq:p}
 p=\mathrm{Pr}[ S > S(\boldsymbol{t}_{1:n} , \boldsymbol{Z}_{1:n}) | H_0 ]= \sum_{j=1}^{2^{n-1}} \mathrm{Pr}[\boldsymbol{Z}_{1:n}=\boldsymbol{z}_{1:n}] p^{(j)},
\end{equation}
which may be expressed in terms of the semi-complete data $p$-values~\eqref{eq:p_i}. A Monte-Carlo approximation of the $p$-value~\eqref{eq:p} may be done,
\begin{equation}\label{eq:pMC}
 p \approx \frac{1}{l}\sum_{j=1}^{l} p^{(a_j)},
\end{equation}
by taking the average of the semi-complete data $p$-values, having their indices in the sampled set $\lbrace a_j \rbrace_{j=1,\dots,l}$.

\section{Monte Carlo Study of the Complete Data EM Algorithm}
\label{sec:caseComplete}

In this section, we discuss the convergence of the EM algorithm (consistency), the power of the goodness of fit test of Sec. \ref{sec:GoF}, and the robustness of Hawkes process estimation in the case of mis-specification of the immigration process. 

\subsection{Parametrization of the Hawkes process with Renewal immigrants}

We consider a particular type of renewal process for immigration, namely a renewal process whose inter-event durations $W$ have a Weibull distribution with conditional intensity
\begin{equation}\label{eq:weib_haz}
	\mu(t)=\frac{\kappa}{\beta}\left(\frac{t}{\beta}\right)^{\kappa-1}, \quad t\ge0~,
\end{equation}
parametrized with the \emph{shape parameter} $\kappa>0$ and \emph{scale parameter} $\beta>0$. In this case, the pdf of inter-event times $W$ is given by:
\begin{equation}\label{eq:weib_pdf}
	g(w)=\mu(w) \exp \left(-\int_{0}^{w}\mu(s)ds \right)
	=\frac{\kappa}{\beta}\left(\frac{w}{\beta}\right)^{\kappa-1}\exp \left( - \left(\frac{w}{\beta}\right)^{\kappa}\right).
\end{equation}

The case $\kappa=1$, for which the intensity~\eqref{eq:weib_haz} is constant, corresponds to the standard Hawkes process~\eqref{eq:HawkesCondInt} with $\mu=1/\beta$. When $\kappa<1$, the intensity decays implying that the inter-event time density~\eqref{eq:weib_pdf} is sub-exponential. In this case, the immigration process is over-dispersed in comparison with the respective Poisson process for $\kappa=1$. On the other hand, as $\kappa\to \infty$, the inter-event density~\eqref{eq:weib_pdf} weakly converges to a delta-function $g(w)=\delta(w-\beta)$, and the immigration process becomes deterministic with regular event spacing $\beta$.
\\
For the offspring density, $h(t)$, we consider both the exponential pdf, originally suggested by Hawkes~\citep{Hawkes1971_orig}:
\begin{equation}\label{eq:exp}
	h_{exp}(t)=\frac{1}{\tau_{0}}\exp\left(-\frac{t}{\tau_{0}}\right)\mathrm{1}_{t\ge0},
\end{equation}
with shape parameter $\tau_{0}>0$; and the Omori-type heavy-tailed pdf \citep{Ogata1988}:
\begin{equation}\label{eq:pow}
	h_{Omori}(t)=\frac{\alpha c^\alpha}{(t+c)^{1+\alpha}}\mathrm{1}_{t\ge0},
\end{equation}
with shift parameter $c>0$ and Pareto tail index $\alpha>0$. 
\\
The exponential offspring density~\eqref{eq:exp}, which is typical for financial and econometric applications~\citep{Bowsher2007,Bauwens2009,FilimonovSornette2012_Reflexivity,Embrechts2011,AitSahalia2011}, ensures Markovian properties to the model~\citep{Oakes1975}. With respect to calibration, it reduces the computational complexity of evaluation of the log-likelihood from $O(n^2)$ to $O(n)$ by taking advantage of a recursive relation~\citep{Ozaki1979} and is more robust to outliers than heavy tailed alternatives~\citep{FilimonovSornette2013_apparent}. A heavy-tailed offspring density~\eqref{eq:pow} is typical for seismological applications~\citep{Ogata2013}, where it accounts for the power law decay of aftershock activity with time (\emph{Omori's law}) in the so-called \emph{Epidemic-Type Aftershock Sequence (ETAS)} models. While the computational complexity of evaluation of the log-likelihood is $O(n^2)$ and cannot be reduced, in practical applications, one can approximate the power law~\eqref{eq:pow} with a sum of weighted exponential functions~\citep{HardimanBouchaud2013}.

\subsection{Starting Points, Convergence, \& Nonparametrics}

We now discuss the sensitivity to starting values, the speed of convergence, and the use of non-parametric estimation in the complete data EM algorithm. The EM algorithm is iterative and requires starting parameter estimates. Further, EM algorithms in general can get stuck in local optima and have a linear rate of convergence~\citep{Dempster1977}. Thus, selecting reasonable starting points, and understanding the speed of convergence are important. However, it was shown in~\cite{Mohler2011} that the EM algorithm for the standard Hawkes process is a projected gradient descent algorithm with superlinear convergence. But, the speed of convergence worsens (towards linear convergence) as clusters are increasingly overlapping. This is intuitively clear, as the branching structure becomes less obvious as clusters overlap. This was explored in detail in~\cite{SornetteUtkin2009}. A comprehensive study of this phenomenon will not be done for this extended algorithm as new insights are not expected. Instead a couple of illustrative examples are given:
\\
{\bf Example I}: A realization of 1000 points of a standard Hawkes process~\eqref{eq:HawkesCondInt} ($\mu=1,~\eta=0.8,~h=0.2 \text{exp}(-t/5)$) with high branching ratio was simulated. Then the Hawkes model with renewal immigration~\eqref{eq:HawkesRenewalCondInt} was estimated on this data using the complete data EM algorithm. The initial parameter estimates were chosen to be very close to a pure Weibull renewal process with parameters estimated by MLE ($\kappa\approx0.6$, $\beta\approx 0.1$). A small offspring component was included in the initial estimate with uniform density on a large support ($\eta=0.05,~h(t)=0.01\lbrace0 <t\leq 100 \rbrace$). 
\\
{\bf Example II}: A realization of 1000 points of a Hawkes process with Weibull renewal immigration~\eqref{eq:HawkesRenewalCondInt},~\eqref{eq:weib_pdf} $~(\kappa=0.7,~\beta=1,~\eta=0.2,~h=0.2 \text{exp}(-t/5))$ with low branching ratio is simulated. Then the Hawkes model with renewal immigration was estimated on this data using the complete data EM algorithm. The initial parameter estimates included Poissonian immigration, a higher branching ratio, and a uniform offspring density ($\kappa=1,~\beta=2, \eta=0.5,~h(t)=0.01\lbrace0 <t\leq 100 \rbrace$). 
\\
The results are shown in Fig. 2 
with example I in the top row and example II in the bottom row. The estimates of the parameters and the offspring density converge well within 50 iterations, however they do not become completely stable. This is due to the nonparametric density estimation, discussed below. Despite the convergence from poor starting estimates for these different synthetic examples, for analysis of real data the authors recommend taking multiple starting points and keeping the best result.
\\
\begin{figure}[h!]
 \begin{center} 
\centerline{\includegraphics[width=13cm]{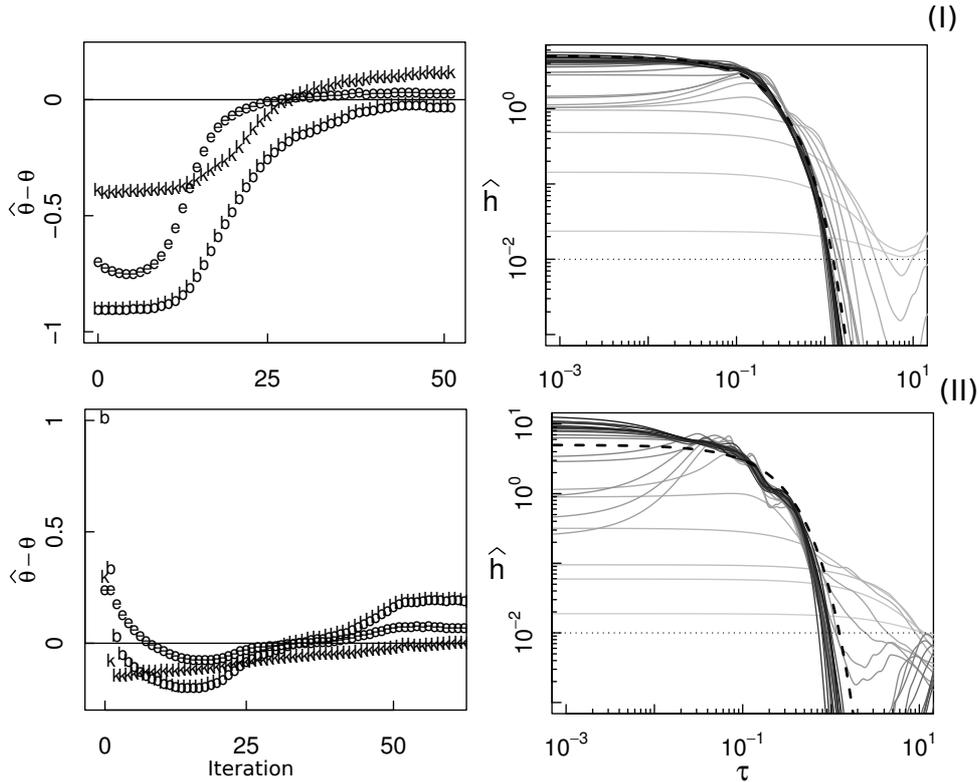}}
\caption{ Convergence of complete data EM estimates of the Hawkes process with renewal immigration for simulated data. Row (I) corresponds to example I, and (II) to example II. The first column provides the difference between the estimated parameter and the true value with parameters $\kappa$, $\beta$, and $\eta$ being represented by k, b, and e respectively. The right column provides the estimated offspring density $\widehat{h}$.  The starting value is the dotted line, and the true function is the bold dashed line. Over iterations, the estimated functions (solid lines) become increasingly dark.   }
\end{center}
\label{fig:convergence}
\end{figure}

The local likelihood non-parametric density estimation technique (implemented in R as locfit) was used \citep{Loader}. This flexible technique estimates the log density using splines in a locally adaptive way. The technique budgets function complexity, only allowing the estimated function to be complex in regions where the data suggests that this is necessary -- avoiding over-fitting and over-smoothing. However it still requires the selection of a smoothing bandwidth parameter. If insufficient smoothness is required, over-fitting will occur. The risk of over-fitting is worsened by the iterative nature of the algorithm. But, the locfit method allows for an effective measure of the degrees of freedom of the estimated function to be computed. With this, one can use, e.g., AIC for the estimated Hawkes model to select the smoothing parameter. Sample weights were considered by using the Monte Carlo method explained in Sec. \ref{sec:compEff}. Thus, from iteration to iteration, the parameter estimates will never become completely stable.
\\
In the examples above, only an exponential offspring density was considered for the data generating process. Heavy tailed (sub-exponential) offspring densities will be difficult to estimate non-parametrically. For instance, most non-parametric density estimation techniques will estimate the tail poorly due to sparse observations. The inclusion of mass in the estimate ``beyond the data'' can be achieved only by requiring a degree of smoothness from the estimated function. However, the assumption of high smoothness will cause problems when estimating regions of the density with high curvature (e.g., a steep mode). If a heavy tailed density is suspected, it may be worth starting with a parametric density such as the Weibull~\eqref{eq:weib_pdf}.


\subsection{Consistency}

This section discusses the bias and efficiency of the Complete Data EM estimator for the Hawkes process with renewal immigration~\eqref{eq:HawkesRenewalCondInt}. 
\\
We have considered synthetic realizations of the process~\eqref{eq:HawkesRenewalCondInt} with parameters $\kappa$ and $\eta$ presented in Table~\eqref{tab:well}. The Weibull shape parameter $\kappa$ was given values $\{0.5,0.75,1,1.25,1.5\}$ ranging from highly over-dispersed to highly under-dispersed. For each value of $\kappa$, the scale parameter $\beta$ was chosen such that the expected immigrant inter-event time was equal to 10, i.e., $\beta$ was given values in $\{5, 8.4, 10, 10.7, 11.1\}$. The characteristic time $\tau_0$ of the exponential offspring density~\eqref{eq:exp} was chosen to be $\tau_0=3$.
\\
For each combination of parameters, 50 independent realizations each of size 500 events were simulated. Efficient simulation was performed using the algorithm of \cite{Moller2005} which exploits the branching process formulation of the Hawkes process. The model parameters $\hat\kappa$, $\hat\beta$, $\hat\eta$, $\hat\tau_0$ were then estimated using the EM algorithm. We intentionally chose ``bad'' starting values for the estimation procedure to demonstrate the convergence of the method: we selected $\hat\kappa^{[0]}=1$; the scale parameter $\hat\beta^{[0]}$ was chosen as the true value $\beta$ multiplied by a uniform random number in $[0.25,4]$; the branching ratio $\hat\eta^{[0]}$ was chosen as a uniform random number in $[0.1,0.9]$; and the characteristic time of the offspring density $\tau_0$ was chosen as a uniform random number in $[0.5,10]$.
\\
The bias and standard deviation of the estimates are presented in Table~\ref{tab:well}. In general, most parameters were well estimated, especially the branching ratio $\eta$.  
Due to the fixed sample size of 500 points, when $\eta$ is larger, the expected number of immigrants $\mbox{E}[N^{(0)}(r)]$ is smaller. Thus the bias and the variance of estimates of immigration process parameters $\hat\kappa$ and $\hat\beta$ are larger with larger $\eta$ and are worst for $\eta=0.9$, i.e., when the $\mbox{E}[N^{(0)}(r)]=50$. Another factor that introduces
systematic error into the results is when $\eta$ is large and thus clusters are overlapping. Intuitively it is clear that when clusters triggered by different immigrant events significantly overlap, it is hard to discern the branching structure. Systematic studies have confirmed that both MLE and EM estimation of the Hawkes process are indeed worse in the case of overlapping clusters~\citep{SornetteUtkin2009,Mohler2011}. However, the bias decreases with increasing sample size.
\\
\begin{table}[t!]
\caption{ Results of the EM estimation (Section~\ref{sec:EM}) of the Hawkes Process with Weibull Renewal immigration~\eqref{eq:HawkesRenewalCondInt} and exponential offspring density on synthetic data.  For each combination of parameters, this table presents the average bias and standard deviation (in brackets) of estimates over 50 synthetic realizations.}\label{tab:well}
\begin{center}
\begin{tabular}{ c c c | c c c c }
\toprule
$\kappa$&$\eta$&$\mbox{E}[N_{(0)}(\tau)]$ &$\widehat{\kappa}-\kappa$&$\widehat{\beta}-\beta$&$\widehat{\eta}-\eta$&$\widehat{\tau_0}-\tau_0$ \\ 
\midrule
0.5 	& 0.1 & 450  &0.02 (0.02) & 0.55 (0.74) & 0.02 (0.06) & -0.05 (2.04) \\	
	& 0.5 & 250  &0.06 (0.03) & 1.70 (1.32)  & 0.03 (0.06) & -0.41 (0.52)  \\
	& 0.9 & 50   &0.16 (0.15) & 0.89 (2.13) & -0.04 (0.05)& -0.46 (0.51)   \\
\midrule	
0.75	& 0.1 & 450  &0.04 (0.04) & 1.01 (1.22) & 0.06 (0.06) & 0.31 (1.51)  \\
	& 0.5 & 250  &0.06 (0.07) & 1.16 (1.52) & 0.02 (0.06) & -0.25 (0.48)  \\
	& 0.9 & 50   &0.12 (0.13) & -1.00 (3.14)& -0.05 (0.05)& -0.52 (0.37)   \\
\midrule	
1   	& 0.1 & 450 &0.02 (0.05) & 0.46 (0.76) & 0.03 (0.04) & 1.71 (2.97)  \\
	& 0.5 & 250 &-0.02 (0.07)& -0.66(1.08) & -0.03 (0.05) & -0.08 (0.48) \\
	& 0.9 & 50  &0.02 (0.15) & -1.58 (2.99)& -0.04 (0.05) & -0.28 (0.60)  \\
\midrule	
1.25	& 0.1 & 450 &-0.03 (0.08) & 0.04 (0.63) & 0.01 (0.04) & 6.23 (6.59)  \\
	& 0.5 & 250 &-0.06 (0.11) & -0.69 (1.21) & -0.04 (0.07)& -0.05 (0.59) \\ 
	& 0.9 & 50  &-0.12 (0.20) & -3.16 (2.53) & -0.07 (0.05) & -0.30 (0.49)  \\
\midrule	
1.5 	& 0.1 & 450 &-0.06 (0.09) & -0.09 (0.6) & 0.00 (0.03) & 3.91 (5.35)    \\
	& 0.5 & 250 &-0.15 (0.10) & -0.79 (1.09) & -0.03 (0.06) & 0.03 (0.56)  \\
	& 0.9 & 50  &-0.29 (0.26) & -3.17 (2.96) & -0.05 (0.05) & -0.36 (0.42)  \\
\bottomrule
 \end{tabular}
\end{center}
 \end{table}

\subsection{Model Selection}

In this section, we address the question of model selection when the immigration process is unknown. For this, we simulate the Hawkes Process with Weibull Renewal immigration~\eqref{eq:HawkesRenewalCondInt} and exponential offspring density~\eqref{eq:exp}, and then test the null hypothesis ($H_0$) that observed events $\{\boldsymbol{t}_{1:n}\}$ are generated with the standard Hawkes model with Poisson immigration~\eqref{eq:HawkesCondInt} versus the alternative hypothesis ($H_1$) that  $\{\boldsymbol{t}_{1:n}\}$ are generated from a Hawkes Process with Weibull Renewal immigration~\eqref{eq:HawkesRenewalCondInt}. In both models ($H_0$ and $H_1$), the offspring density is assumed to be exponential~\eqref{eq:exp}.
\\
We consider three statistical test: (i) comparison of the AIC values of $H_0$ and $H_1$, (ii) the Wilks likelihood ratio test with level 0.05~\citep{Wilks1938} where $H_0$ is nested in $H_1$ and (iii) the KS test of residuals (transformed time events) with level 0.05 for $H_0$ discussed in Section~\ref{sec:GoF}. The option (iii) is not a test of $H_0$ against alternative $H_1$, but the Portmanteau-type test of $H_0$ against the alternative hypothesis $\tilde H_1$, that is loosely specified (i.e. ``not $H_0$'').
\\
The parameters for the process~\eqref{eq:HawkesRenewalCondInt},~\eqref{eq:weib_haz} were chosen as follows: the Weibull shape parameter $\kappa$ was given values $\{0.5,0.75,1,1.25,1.5\}$ and other parameters were not swept, taking values of $\beta=1$, $\eta=0.6$ and $\tau_0=0.3$. For each combinations of these parameters, we have simulated 100 independent realizations of size 250, 500, 750, and 1000 events. Then both models were estimated on each sample: the true model~\eqref{eq:HawkesRenewalCondInt}, which corresponds to $H_1$, is estimated using the complete-data EM algorithm, and the misspecified model~\eqref{eq:HawkesCondInt}, which corresponds to $H_0$, is estimated using straightforward maximization of the log-likelihood~\eqref{eq:lik}. The monte-carlo approximation of the likelihood for the true model~\eqref{eq:LikAggMC} was done with 200 sampled likelihoods.
\\
Table~\ref{tab:power} summarizes the results. In general, the larger the sample, and the further from equidispersed immigration (when $\kappa$ is away from 1), the more powerful the test. AIC provides a very powerful decision rule for comparing the models, even for small sample sizes (e.g., $n=250$) and moderately over and under dispersed immigration (e.g., $\kappa=0.75$ and $\kappa=1.25$ respectively). When the null model is true (i.e., $\kappa=1$), both models provide approximately equal AIC. For this reason, the more complex model should be chosen when the difference in AIC is ``significantly'' greater than zero, otherwise the AIC decision would have a high level. The Wilks test is very powerful in general for sample sizes with 500 or more points, and at even smaller sample sizes given high over or underdispersion (e.g., $\kappa=0.5$ and $\kappa=1.5$ respectively) in the immigration. However, for simulations from the null model (when $\kappa=1$), the Wilks test has too low a level, often rejecting less than 5 percent of the time. This could be because of limited accuracy in the approximation of the likelihood using~\eqref{eq:LikAggMC} with $l=100$ samples, or numerical imprecision when averaging the likelihoods. The KS test is understandably the least powerful as it specifies no alternative model. Even on large sample sizes ($n=1000,~E[N^{(0)}(r)]=400$) and for significant immigrant overdispersion ($\kappa=0.5$), the test has very low power, less than 0.5. 
\\
 
   \begin{table}[h!]
\begin{center}
\begin{tabular}{ c c c | c c c c c }
\toprule
 Test & n    & $E[N^{(0)}(r)]$ &$\kappa=0.5$& $\kappa=0.75$ & $\kappa=1$  & $\kappa=1.25$ & $\kappa=1.5$	\\
\midrule
  AIC  		& 250  & 100 &   0.99	&0.58	&0.06	&0.35	&0.81	\\
		& 500  & 200 &   1	&0.79	&0.07	&0.6	&0.95	\\
		& 750  & 300 &   1	&0.93	&0.12	&0.68	&0.99	\\
		& 1000 & 400 &   1	&0.96	&0.2	&0.83	&1	\\
\midrule
  Wilks		& 250  & 100 &   0.97	&0.35	&0.01	&0.19	&0.54	\\
		& 500  & 200 &   1	&0.67	&0.01	&0.32	&0.87	\\
		& 750  & 300 &   1	&0.85	&0.04	&0.5	&0.95	\\
		& 1000 & 400 &   1	&0.9	&0.05	&0.65	&1	\\
\midrule
  KS     	& 250  & 100 &   0.08	&0.05	&0.03	&0.06	&0.1	\\
		& 500  & 200 &   0.17	&0.04	&0.03	&0.08	&0.17	\\ 
		& 750  & 300 &   0.30	&0.04	&0.06	&0.13	&0.22	\\ 
		& 1000 & 400 &   0.46	&0.06	&0.05	&0.14	&0.22	\\ 
\bottomrule
\end{tabular}
 \caption{ Results of model selection tests. $\mbox{E}[N^{(0)}(r)]$ denotes the expected number of immigrant events in the sample. AIC provides the fraction of the 50 repetitions in which the $H_{1}$ model had superior AIC to the $H_{0}$ model. Wilks provides the fraction of the 50 repetitions in which the $H_{0}$ model was rejected when compared to the $H_{1}$ model using the Wilks test at level 0.05. KS provides the fraction of the 50 repetitions in which the $H_{0}$ model was rejected when using the KS test at level 0.05.
 }\label{tab:power}
\end{center}
 \end{table}
 
  Following from the analysis presented above, model selection can be successfully resolved using AIC and/or the Wilks test. In the following section, we will see that misspecification of the model (misspecification of the immigration process) can significantly bias parameter estimates.

\subsection{Robustness of Branching Ratio Estimation under Mis-specification of the Immigration Process}\label{sec:robust}
 
In this section, we explore the robustness of the estimation of the branching ratio~\eqref{eq:etaEM} when the immigrant process is mis-specified in the Hawkes model. For this, we consider again the Hawkes Process with Weibull Renewal immigration~\eqref{eq:HawkesRenewalCondInt},\eqref{eq:weib_haz} and exponential offspring density~\eqref{eq:exp}. We fixed parameters $\eta=0.5$ and $\tau=0.1$ and varied the immigration shape parameter $\kappa$ from highly over-dispersed 0.4 to under-dispersed 1.4. As before, the scale parameter $\beta$ was chosen so that, for any given $\kappa$, the expected immigrant inter-event time was fixed (in this case $\mbox{E}[T^{(0)}_{i}-T^{(0)}_{i-1}]=4$). 
\\
For each value of $\kappa$, we have simulated 50 independent realizations. Each realization was used for parametric MLE of the standard Hawkes process with Poisson immigration~\eqref{eq:HawkesCondInt} and (i) exponential offspring density~\eqref{eq:exp} and (ii) Omori-type density~\eqref{eq:pow}. 
Figure 3 
presents results of the estimation of the branching ratio $\hat\eta$ as a function of the shape parameter $\kappa$ of the underlying immigration process. 
\\
\begin{figure}[h!]
 \begin{center} 
\centerline{\includegraphics[width=7cm]{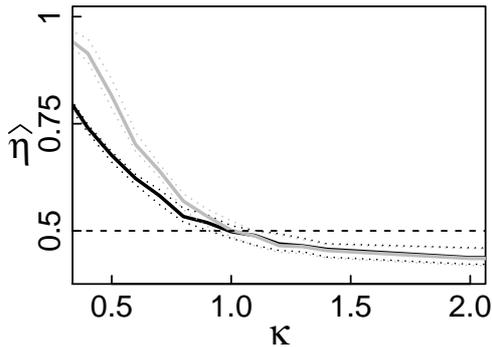}}
\caption{Estimates of the branching ratio $\hat\eta$ \eqref{eq:etaEM} using the Hawkes model with Poissonian immigration and exponential offspring density (black) and Omori-type density (grey) when the true process is generated with the Hawkes model with Weibull Renewal immigration with shape parameter $\kappa$. The true branching ratio (0.5) is presented with a horizontal dashed line. Solid lines correspond to median values and dotted lines present quartiles of estimates for both kernels.}
\end{center}
\label{fig:varying_k}
\end{figure}

As seen from Figure 3, 
both models with Poisson immigration have a significant bias in the estimation of $\hat\eta$. In the case of under-dispersed immigration ($\kappa>1$), one observes a relatively small negative bias which is similar for exponential~\eqref{eq:exp} and Omori-type~\eqref{eq:pow} offspring densities. In contrast, for over-dispersed immigration ($\kappa<1$) the bias is positive and much stronger. For instance, when $\kappa=0.5$, the branching ratio has median positive bias of 0.17 and 0.31 for the Hawkes process with exponential and Omori-type offspring densities respectively.

\section{Monte Carlo Study of The Semi-Complete Data EM Algorithm}
\label{sec:study2}

In this section, a Monte Carlo study is done, using the semi-complete-data EM algorithm to estimate the standard Hawkes process~\eqref{eq:HawkesCondInt} with deterministic immigration intensity $0<\mu(t)<\infty,~\forall t$. Of course, this algorithm may also estimate the Hawkes process with Renewal process immigration. For computational efficiency, an exponential offspring density~\eqref{eq:exp} is chosen. In this case, both the E and M steps of the semi-complete-data EM algorithm are $O(n)$ and it thus becomes possible to estimate the model on large datasets with a standard PC (e.g., estimation on a sample of tens of thousands of points takes a few minutes). The immigration intensity will be estimated using kernel estimation  
\begin{equation}\label{eq:semiMu}
\widehat{\mu(t)}= \sum_{i=1}^{n} \pi_{i} k(t-t_{i};b) 1_{\lbrace 0<t<r \rbrace}  + c(t)~,
\end{equation}
where the kernel function $k(t;b)$ is a pdf with bandwidth parameter $b$. This estimator~\eqref{eq:semiMu} distributes mass $\pi_{i}$ around each point $t_{i}$. The higher the bandwidth, the more dispersed the mass is. One practical issue is that mass may be distributed outside of the interval $(0,r]$. This may easily be solved by symetrically ``reflecting'' any mass outside of the interval back into the interval. This operation is denoted by the term $c(t)\geq 0,~0<t\leq r$. Another practical issue is the selection of a kernel density and the bandwidth. In general, this involves managing the trade off between model complexity and goodness of fit. For more detail on how to select the bandwidth, see~\cite{Silverman1986,Turlach1993}. A nice feature of this estimator~\eqref{eq:semiMu}) is that $\int_{0}^{r} \widehat{\mu(t)}=\sum_{i=1}^{n}\pi_i$, i.e., it is an unbiased estimator for the total number of immigrant points in the sample.  An important consequence of this is that there is not systematic error in the estimation of the branching ratio~\eqref{eq:etaEM}.
\\
\begin{figure}[h!]
 \begin{center} 
\centerline{\includegraphics[width=8cm]{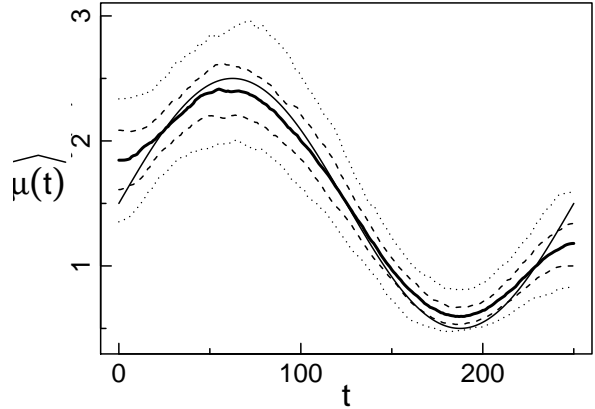}}
\caption{ 
The solid thin black line is the true sinusoidal immigration intensity used in simulation. Lines are also plotted for the median (heavy solid), quartiles (dashed), and 0.05 and 0.95 quantiles (dotted) of all estimates.
}
\end{center}
\label{fig:inhomo_mu}
\end{figure}

This Monte Carlo study involves simulating the Hawkes process with sinusoidal immigration intensity $\mu(t)=\sin( 2\pi t/250 )+1.5$, exponential offspring density~\eqref{eq:exp}
with scale parameter $\tau_0=0.1$, and branching ratio values $\eta$ sweeping 0.1 to 0.9 by 0.1. For each set of parameters, 50 simulations of this process on one period of the immigration intensity $(0,250]$ were performed. The median sample size was 1200 with quartiles 520 and 1310. Two models were estimated on each realization using the Semi-Complete Data EM algorithm (see section \ref{sec:semiEM}): the first being the true model, and the second (the false model) being the true model but with homogeneous immigration (i.e., $\mu(t)=\mu$). The initial parameter estimates were randomly chosen uniformly at random in the following intervals $\eta \in (0.1,0.9)$, $\tau_0 \in (0.1,10)$ and $\mu\in (0.1,5)$. The EM algorithm was allowed to perform 200 iterations, but in 90 percent of the time converged in less than 100 iterations. The convergence criterion was that the cumulative sum of the absolute differences of estimated parameters for the previous 3 iterations were within $10^{-6}$.
\\
In Fig.~4, 
the true immigration intensity and a summary of the estimated immigration intensity across all samples is given for the estimated true model. The immigration intensity is well estimated, including cases when most of the points are offspring (when $\eta=0.9$).
\\
In Fig.~5, 
the errors in the estimation of the branching ratio for the true model (grey) and the false model (black) are summarized. For the true model, estimation is consistent and efficient. For the false model, the branching ratio is consistently overestimated, in particular for low values of the branching ratio -- i.e., an upward bias of more than 0.6 when the true branching ratio $\eta=0.1$, and still an upward bias of approximately 0.1 when $\eta=0.5$. The overestimation is due to the fact that the inter-event time process of the inhomogeneous immigration is over-dispersed and contains apparent clustering. Estimation improves with increasing $\eta$ because, as the number of offspring points increases, the memory of the estimated offspring becomes concentrated on the shorter offspring timescale. As a consequence, the dispersion / clustering in the immigration process can no longer be attributed to the memory kernel.

\begin{figure}[h!]
 \begin{center} 
\centerline{\includegraphics[width=8cm]{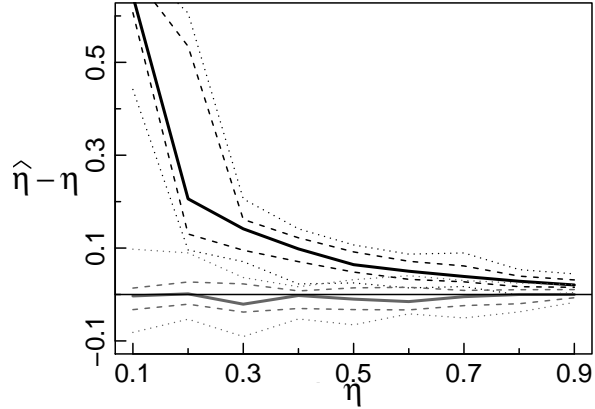}}
\caption{ 
The difference between the estimated branching ratio and the true branching ratio from repeated simulation and estimation. The horizontal axis is the value of the true branching ratio. The grey lines are for the true model and the black lines are for the false model. The median (heavy solid), quartiles (dashed), and 0.05 and 0.95 quantiles (dotted) of all estimates are given.
}
\end{center}
\label{fig:inhomo_eta}
\end{figure}

\section{Discussion}

In this article, the Hawkes process with renewal immigration~\eqref{eq:HawkesRenewalCondInt} was proposed and estimation was made possible by the introduction of two expectation-maximisation (EM) algorithms, the first being an extension of \cite{Veen2008}. These estimation techniques were shown to be consistent in simulation studies, and easily allow for non-parametric estimation (more easily than in \cite{LewisMohler2010}). Further, a computationally efficient implementation of the semi-complete data EM algorithm was shown to be useful for estimating an inhomogeneous Poissonian background intensity. The importance of correctly specifying the immigration process on branching ratio estimation was highlighted -- indicating strong potential for the overestimation of the branching ratio when poissonian immigration is falsely assumed.
\\
As has been discussed, the Hawkes model has been used in many areas. We recommend as a best practice that the Hawkes process with renewal immigration be considered as an alternative model in such studies, in particular, when quantification of the branching ratio is of interest and one observes highly dispersed locations of clusters in the data. Alternatively, when one observes that the rate of events is changing, the inhomogeneous specification of the immigration process should be chosen. 
\\
Let us finish by providing a few examples of the relevance of the results of this paper to the existing literature. Attempts to quantify the branching ratio for high frequency prices fluctuations of the S\&P 500 e-mini futures contracts where the rate of events changes significantly over the day~\citep{FilimonovSornette2012_Reflexivity,HardimanBouchaud2013} can profit from using the semi-complete data EM algorithm with non-parametric estimation of the inhomogeneous Poissonian background intensity. Further, in~\cite{FilimonovSornette2013_apparent}, it was shown that the empirical inter-event time distribution of the S\&P 500 e-mini futures contracts price fluctuations has too heavy of a tail to be explained by a Hawkes process with Poissonian immigration. Allowing for overdispersed renewal immigration could be helpful here. Either of these approaches should lower the estimated branching ratio from the levels reported in~\cite{HardimanBouchaud2013}. For the modeling of rainfall, this approach provides a richer model than \cite{Cowpertwait2000} and superior estimation to \cite{Salim2003}.
\\
There are a number of further methodological research directions suggested by this paper. In general, the EM algorithm enables the maximum likelihood estimation of multi-type point processes, where the type is unobserved; for instance, in other types of cluster processes, potentially with renewal process immigration. Within the Hawkes model, the EM algorithm may also be extended to the case of the marked Hawkes model -- in which the size of an event influences its expected number of offspring via its fertility function -- allowing for easy non parametric estimation of the fertility function. Additionally, the Hawkes model could be further extended to have self-exciting immigration, that is, a model with two stages of clustering.


\bibliographystyle{plainnat}

\end{document}